\documentclass[12pt%
]{article}
  
\usepackage[scaled=0.92]{helvet} 
\usepackage{setspace} 
\usepackage{latexsym}     
\usepackage{amssymb,amsmath, bm}
\usepackage{graphicx} 
\usepackage{marvosym}
\usepackage{multirow}
\usepackage{xcolor,colortbl}
\usepackage[title]{appendix}
 
\usepackage{tikz}
\usetikzlibrary{shapes,arrows}
\usetikzlibrary{positioning}

\usepackage{subcaption}
 
\usepackage{natbib} 
\usepackage{color}
\usepackage[hidelinks, pdftex, bookmarksopen=true, bookmarksnumbered=true,
pdfstartview=FitH, breaklinks=true, urlbordercolor={0 1 0}, citebordercolor={0 0 1}]{hyperref} 
\definecolor{darkgreen}{rgb}{0,0.545,0}
\definecolor{darkyellow}{rgb}{0.933,0.604,0}

\usepackage{longtable}
\usepackage{booktabs}
\usepackage{tabularx}

\usepackage{dcolumn}
\newcolumntype{.}{D{.}{.}{-1}}
\newcolumntype{d}[1]{D{.}{.}{#1}}

\usepackage{rotating}
\usepackage{bbm}

\newcommand{\blind}{0}

\usepackage{theorem}
\theoremstyle{break}
\theoremheaderfont{\scshape}
\theorembodyfont{\rm}

\newcommand\addprime{^{\prime}}
\newcommand\addtop{^{\top}}
\newcommand\addast{^{\ast}}

\newcommand\iid{\stackrel{\sf i.i.d.}{\sim}}

\renewcommand\r{\right}
\renewcommand\l{\left}

\newcommand\E{\mathbb{E}}

\usepackage{arydshln} 
 
\usepackage[compact]{titlesec}

\allowdisplaybreaks[4]
\graphicspath{{figs/}}

\begin{document}

\newcommand\spacingset[1]{\renewcommand{\baselinestretch}% 
{#1}\small\normalsize}

\spacingset{1.1}

%\makeatletter
%\def\thefootnote{\ifnum\c@footnote>\z@\leavevmode\lower.5ex\hbox{$^{*}$}\fi}
%\makeatother

\newcommand{\tit}{\textbf{%
A Non-parametric Bayesian Model for Detecting Differential Item Functioning: An Application to Political Representation in the US%
}}

\if0\blind
{
 \title{\tit\thanks{
 \protect We would like to thank participants in the 2019 Asian PolMeth conference and at the UCLA Political Science Methods Workshop for their useful feedback.
 }
 } 

 \author{
 Yuki Shiraito\thanks{
    Assistant Professor, Department of Political Science, University of Michigan. Center for Political Studies, 4259 Institute for Social Research, 426 Thompson Street, Ann Arbor, MI 48104-2321.  Phone: 734-615-5165, Email: \href{mailto:shiraito@umich.edu}{\tt shiraito@umich.edu}, URL: \href{https://shiraito.github.io}{\tt shiraito.github.io}.
    }
    \and
 James Lo\thanks{
    Assistant Professor, Department of Political Science and International Relations, University of Southern California, 3518 Trousdale Parkway, CPA 327, Los Angeles, CA, 90089. Email: \href{mailto:lojames@usc.edu}{\tt lojames@usc.edu}
    }
    \and
 Santiago Olivella\thanks{
    Associate Professor, Department of Political Science, University of North Carolina at Chapel Hill, Chapel Hill NC. 361 Hamilton Hall, CB 3265, Chapel Hill, NC 27599. Email: \href{mailto:olivella@unc.edu}{\tt{olivella@unc.edu}}, URL: {\href{https://santiagoolivella.info}{\tt santiagoolivella.info}}
    }
 }

 \date{
 First draft: December 18, 2018 \\
 This draft: September 6, 2022 
 }
}\fi
 
\if1\blind
{
 \title{\bf \tit}
 \date{}
}\fi

\maketitle

\pdfbookmark[1]{Title Page}{Title Page}

\thispagestyle{empty} 
\setcounter{page}{0}   

\begin{abstract}
A common approach when studying the quality of representation involves comparing the latent preferences of voters and legislators, commonly obtained by fitting an item-response theory (IRT) model to a common set of stimuli. Despite being exposed to the same stimuli, voters and legislators may not share a common understanding of how these stimuli map onto their latent preferences, leading to differential item-functioning (DIF) and incomparability of estimates. We explore the presence of DIF and incomparability of latent preferences obtained through IRT models by re-analyzing an influential survey data set, where survey respondents expressed their preferences on roll call votes that U.S. legislators had previously voted on. To do so, we propose defining a Dirichlet Process prior over item-response functions in standard IRT models. In contrast to typical multi-step approaches to detecting DIF, our strategy allows researchers to fit a single model, automatically identifying incomparable sub-groups with different mappings from latent traits onto observed responses. We find that although there is a group of voters whose estimated positions can be safely compared to those of legislators, a sizeable share of surveyed voters understand stimuli in fundamentally different ways. Ignoring these issues can lead to incorrect conclusions about the quality of representation.

\vspace*{12pt}

\end{abstract}

\clearpage

%=== footnote number in the style of #)
% \makeatletter
% \def\thefootnote{\ifnum\c@footnote>\z@\leavevmode\lower.5ex\hbox{$^{\@arabic\c@footnote)}$}\fi}
% \makeatother

\doublespacing

\section{Introduction}
Measurement models, such as the popular two-parameter Item Response Theory (IRT) Model, are commonly used to measure latent social-scientific constructs like political ideology. Such models use observed responses to a common set of stimuli (e.g., congressional bills to be voted on) in order to estimate underlying traits of respondents and mappings from those traits to the responses given (e.g., a `yea' or `nay' vote).
Standard applications of these models typically proceed on the assumption that the set of stimuli used to measure constructs of interest are understood equally by all respondents, thus making their answers (and anything we learn from them) comparable. This assumption is commonly known as \emph{measurement invariance}, or \emph{measurement equivalence} \citep{King_etal2004,stegmueller2011}.

As early as 1980, however, researchers were aware that violations of this assumption were possible.
%For example, in the Congressional example, it may be the case that female legislators respond to common stimuli in different ways from their male counterparts even if their underlying latent traits were the same.
Today, violations of this assumption are commonly referred to as Differential Item Functioning (DIF). In the language of the time, \citet[pg. 212]{lord2012applications} defined DIF by stating that ``if an item has a different item response function for one group than for another, it is clear that the item is biased.''

Since Lord's description of the problem that DIF poses to measurement, a number of researchers have developed and adopted various techniques to mitigate its effects. \citet{lord2012applications,lord1977study} proposed a general test of joint difference between the item parameters estimates for two groups of respondents in the data. \citet{thissen1993detection} build on this work, proposing additional methods for fitting IRT models to a known reference and focal group and then testing for the statistical differences in item parmeters between the two groups. This work in \emph{identifying} DIF is complemented by work that attempts to \emph{rescale} DIF under very specific circumstances and assumptions, including \citet{King_etal2004,stegmueller2011,Aldrich_McKelvey1977,poole1998recovering,hare2015using,jesseeestimating}.

%Measurement invariance is violated when respondents in different sub-groups (e.g., political elites vs. non-elites) understand stimuli differently. In such instances, members of different groups effectively use different mental mappings from their traits onto the answers they give. Accordingly, assuming that these mappings are the same for all respondents when they are, in fact, different, can result in biased estimates of the latent constructs the model sets out to uncover. Unaddressed violations of measurement invariance --- commonly known as \emph{differential item functioning} --- may thus preclude meaningful interpretation of the estimated latent constructs \citep{King_etal2004}.

In this paper, we propose a model designed to improve measurement when DIF is present. To do so, we rely on Bayesian non-parametrics to flexibly estimate differences in the mappings used by respondents when presented with a common set of items. While we are not the first scholars to combine Bayesian non-parametric techniques (and specifically the Dirichlet process) with IRT models \citep[see, for example,][]{miyazaki2009bayesian,jara2011dppackage}, to the best of our knowledge, we are the first to do so explicitly with the goal of addressing differential item functioning. Our model --- which we refer to as the Multiple Policy Space (MPS) model --- addresses one specific violation of measurement invariance that is of particular importance in political methodology.

Our model identifies sub-groups of respondents who share common item parameter values, and whose positions in a shared latent space can thus safely be compared. Thus, while sub-groups in our model will not necessarily be distinct from each other, our model can estimate group-specific latent traits by first learning a sorting of observations across unobserved groups of respondents who share a common understanding of items, and conditioning on these group memberships to carry out the measurement exercise. This is similar in spirit to work done by \citet{lord2012applications} and \citet{thissen1993detection}, but a crucial difference in our work is that we do not require researchers to \emph{a priori} specify a set of group memberships of members before testing. Rather, our work offers an automated, model-based approach to discover these group memberships from response patterns alone, which in turn also identifies groups of respondents for who common latent trait mappings can and cannot be validly compared. In discovering these latent group memberships, we can also distinguish the set of respondents in our data that are comparable on a common latent score (i.e., a liberal-conservative ideological spectrum) from those who think on a different dimension (i.e., a libertarian-authoritarian spectrum).\footnote{%
\label{fn:samesubstantivecluster}
Ideal points of people belonging to the same substantive cluster are comparable, assuming that we take the spatial model of voting as our preferred model of political preferences. While it is possible to compare preferences on individual issues of individuals in separate clusters (i.e., opinions on tax cuts), we see no straightforward way to standardize ideal points of individuals in different ideological clusters (i.e., a 0.5 on a liberal-conservative scale vs a -0.5 on a libertarian-authoritarian scale).}

To empirically illustrate our model, we apply it to the estimation of political ideology using a data set that contains both legislators and voters. Our application is based on the data set analyzed by \citet{jessee2016can}, which contains 32,800 respondents in a survey conducted in 2008 and 550 U.S.~Congress members who served in the same year. As we discussed above and will elaborate in the next section, the aim of the MPS model in this application is to identify subsets of the voters and legislators within which item response functions are shared and to measure latent traits within each subset, rather than jointly scaling the actors into a common ideology space or determining whether joint scaling disrupts ideal point estimates or not. In our analysis, we find that the 73\% of the voters in the data set share item parameters with the legislators, whereas the 27\% of the voters do not.

Our paper proceeds as follows.
First, we introduce the substantive context and data set of our application, focusing on the work of \citet{jessee2016can}.
Second, we discuss and motivate the details of our IRT model for dealing with measurement heterogeneity, discussing the role of the Dirichlet Process prior---the underlying technology that our proposed model uses to non-parametrically separate respondents into groups. Third, we offer Monte Carlo simulation evidence demonstrating the ability of our model to recover the key parameters of interest. Fourth, we present a substantive application of our model to the debate on the joint scaling of legislators and voters. This debate focuses on the extent to which we can reasonably scale legislators and voters into the same ideological space, which effectively can be re-framed as a question regarding the extent to which voters share the same item parameters as legislators. We conclude with some thoughts on potential applications of our approach to dealing with heterogeneity in measurement.

\section{Application: Scaling Legislators and Voters}
%%
%% Yuki's note for self
%% Move the data and problem description here
In recent years, a literature extending the canonical two-parameter IRT model to jointly scale legislators and voters using bridging items has emerged \citep{bafumi2010leapfrog,jessee2012ideology,hirano2011policy, saiegh2015}. In such applications, researchers begin with a set of items that legislators have already provided responses to, such as a set of pre-existing roll call votes. Voters on a survey are then provided with the same items and asked for their responses. The responses of the voters and legislators are grouped together and jointly scaled into a common space, providing estimated ideal points of voters and legislators that in theory can then be compared to one another.

In an influential critique of this work, \citet{jessee2016can} argued that this approach did not necessarily guarantee that legislators and voters could jointly be scaled into a common space.\footnote{A critique of joint scaling by \citet{lewis2013} is conceptually similar to Jessee's critique in sharing concern that parameter values for different groups of respondents differ, but employs a different methodology. } Jessee's core critique was that legislators and voters potentially saw the items and the ideological space differently, even if they were expressing preferences on the same items. Joint scaling effectively constrains the item parameters for those items to be identical for both groups, but does not guarantee that they are actually identical in reality. In the language of the MPS model, Jessee claimed that there were potentially two separate clusters --- one for legislators and another for voters --- through which differential item functioning can occur. 

For Jessee, the question of whether voters and legislators could be jointly scaled was essentially a question of sensitivity analysis. He conceptualized the answer to this question as a binary one --- that is, either all voters and legislators could be jointly scaled together, or they could not be. His proposed solution to answer this question was to separately estimate two separate models for legislators and voters. Jessee then used the legislator item parameters to scale voters in ``legislator space'', and the voter item parameters to scale legislators into ``voter space''. If these estimates were similar to those obtained via joint scaling, then the results were robust and legislators and voters could be scaled together. The Jessee approach essentially adopts \citet{lord2012applications} and \citet{thissen1993detection} approach for testing for DIF, and adds an extra step by re-estimating latent traits for the reference and focal groups conditional on the item parameters of the other group.

Our approach to answering this question differs substantially from Jessee, but it is worth noting that his conception of the problem is a special case of our approach. To answer this question using our model, we can estimate an MPS model where we constrain all of the legislators to share a common set of item parameters, but allow voters to move between clusters. Voters can thus be estimated to share membership in the legislator cluster, or they can split off into other separate clusters occupied only by voters. This highlights the principal difference between the MPS model and Jessee's approach. Jessee's approach is a sensitivity analysis in the spirit of Lord (1980) that provides a binary Yes/No answer to the question of whether jointly scaling legislators and voters together will change the ideal points estimates meaningfully --- that is, it scales voters using the item parameters of the legislators, and legislators using the item parameters of the voters. Substantial deviation in the estimated ideal points between these approaches suggests that voters and legislators cannot be scaled together in a common space. In contrast, the MPS model identifies the subset of voters that can be jointly scaled with legislators, which the Jessee model does not. While two special cases of the MPS model (i.e., either all voters lie share item parameters with the legislators, or none of them do) correspond to potential answers that Jessee's model can provide, our model can provide intermediate answers --- notably, we can identify the number and identity of the voters who share an ideological space with legislators, and voters need not all share a common ideological space with one another.

\section{Model Description}
%\subsection*{Group-Based Differential Item Functioning}
Our modeling approach adopts the same group-based definition of differential item functioning previously described by \citet{lord2012applications} and \citet{thissen1993detection}. Specifically, we assume that there are subsets of respondents who share the same item response functions, which in turn are different from those used by members of other subsets. 

If we knew \emph{a priori} what these groups were (e.g., gender of legislators in legislative voting), correcting/accounting for differential item functioning would be relatively easy, and would amount to conditioning on group membership during the scaling exercise. However, the subsets of respondents for whom items are expected to function in different ways is often not immediately obvious. In such cases, we can use response patterns across items to \emph{estimate} membership into groups of respondents defined by clusters of item parameter values (i.e., of the parameters that define different item response functions). This is the key insight behind our approach, which relies on a Dirichlet process prior for item parameters that allows us to identify collections of individuals for whom IRFs operate similarly without the need to fix memberships or the number of such groups \emph{a priori}.     

 To this end, we propose a model that addresses DIF violations occurring across groups of respondents.  When group membership is held constant across items, we are able to identify sets of respondents who are effectively mapped onto different spaces, but who are guaranteed to be comparable \emph{within} group assignment. Our approach, which we call the \emph{Multiple Policy Space} (MPS) model, is a latent-variable generalization of the standard non-parametric Dirichlet process mixture regression model \citep[e.g.][]{hannah2011dirichlet}.\footnote{As such, it differs from other uses of the DP prior (DPP), such as that of \citet{kyung2009} or \citet{traunmuller2015}, where a DPP is defined as part of a semi-parametric model.}   

With these intuitions in place, we now present our DP-enhanced IRT model, including a discussion of how the Dirichlet Process prior can help us address the issue of heterogeneous item response functions, but leave the details of our Bayesian simulation algorithm to the appendix.   

\section*{The Multiple Policy Space Model}
Let $y_{i,j}\in\{0,1\}$ be respondent $i$'s ($i\in{1,\ldots,N}$) response on item $j\in{1,\ldots,J}$. Our 2-parameter IRT model defines
\begin{align}
\begin{split}
  y_{i,j} \mid \bm{\theta},\bm{\beta},\gamma &\iid \mathcal{B} \l( \Phi \l( \bm{\beta}_{k[i],j}\addtop \bm{\theta}_{i} - \gamma_{k[i],j} \r) \r),\; \forall i,j \\
  \bm{\theta}_i &\iid \mathcal{N}_{D} \l( \bm{0},\bm{\Lambda}^{-1} \r), \; \forall i\\
 (\bm{\beta}_{k,j},\gamma_{k})&\iid \mathcal{N}_{D+1} \l( \bm{0},\bm{\Omega}^{-1} \r),\;  \forall k,j
 \end{split}
 \label{eq:irt}
\end{align}
 where $k[i] \in 1, \ldots$ is a latent cluster to which respondent $i$ belongs; $\bm{\theta}_i$ is a vector of latent
 respondent positions on $D$-dimensional space;
 $\bm{\beta}_{k,j}$ is a vector of cluster-specific
 item-discrimination parameters; $\gamma_{k,j}$ is a
 cluster-specific item-difficulty parameter.\footnote{$\bm{\Lambda}$ and $\bm{\Omega}$ are prior precisions of ideal points and item parameters, respectively, with $\bm{\Lambda}\equiv\mathbf{I}_D$ for identification purposes.} Substantively, cluster-specific item parameters reflect the possibility that the IRF is shared by respondents belonging to the same group $k$ but heterogeneous across groups. 
 
 To aid in the substantive interpretation of this model, it is helpful to consider the case where we only keep respondents in group $k = k'$, and discard respondents belonging to all other groups. Thus, we are only using the item parameters from the cluster $k'$, which are common to all respondents in that cluster. Since this is the case, we can discard the cluster indexing altogether, and the first line of Equation~\eqref{eq:irt} reduces to:
\begin{align*}
  y_{i,j} \mid \bm{\theta},\bm{\beta},\gamma &\iid \mathcal{B} \l( \Phi \l( \bm{\beta}_{j}\addtop \bm{\theta}_{i} - \gamma_{j} \r) \r),\forall ~i\text{ s.t. } k[i] = k\addprime 
\end{align*}

This is the standard two-parameter IRT model. Thus, we can summarize our model as follows: if cluster memberships were known, the MPS model is equivalent to taking subsets of respondents by cluster, and scaling each cluster separately using the standard two-parameter IRT model. This implies that even though they are expressing preferences on the same items, respondents in different clusters are mapping the same items onto different latent spaces. Thus, comparisons of $\bm{\theta}_i$ are only meaningful when those $\bm{\theta}_i$ belong to the same cluster (i.e., would have been scaled together in the same IRT model).\footnote{Item parameters follow a similar logic in the sense that they are only comparable within the same cluster, but not across clusters.}

 Given that we do not observe which observations belong to which clusters, however, we need to define a probabilistic model for the cluster memberships that does not require \emph{a priori} specifying how many clusters respondents can be sorted into. For this, we rely on the Dirichlet Process prior.

\subsection{Sampling cluster memberships using a Dirichlet Process mixture}

The Dirichlet process is a popular non-parametric Bayesian prior (\citealt{ferguson1973bayesian}.  See also \citealt{teh2010dirichlet}).
The basic idea of the Dirichlet process is that any sample of data for which one typically estimates a set of parameters can be split into subgroups of units, letting the data guide discovery of those groups instead of requiring users to pre-specify their number \emph{a priori}. Technically, the Dirichlet process prior allows mixture models to have a potentially infinite number of mixture components, but in general it allows a small number of components to be occupied by observations by penalizing the total number of occupied components. It is known that the number of mixture components is not consistently estimated. Nevertheless, when used for density estimation \citep{ghosal1999posterior} and non-parametric generalized (mixed) linear models \citep{hannah2011dirichlet, kyung2009}, Dirichlet process mixture models consistently estimate the density and the mean function, respectively.

We now describe the Dirichlet process mixture of our multiple policy space model.\footnote{The description of the Dirichlet process here is based on the stick-breaking construction developed by \citet{sethuraman1994constructive}.} Let $p_{k^\prime}$ denote the probability that each observation is assigned to cluster $k^\prime$, for $k^\prime = 1, 2, \dots$, i.e., $p_{k^\prime} \equiv {\rm Pr}(k[i] = k^\prime)$, and let the last line of Equation~\eqref{eq:irt} be the base distribution from which cluster-specific item parameters are drawn. Then under a DP-mixture model of cluster-specific IRT likelihoods, we have
\begin{align}
 k[i] & \iid {\rm Categorical} \left( \left\{ p_{k^\prime} \right\}_{k^\prime = 1}^{\infty} \right) \label{eq:clusterassign}\\
  p_{k^\prime} &= \pi_{k^\prime} \prod_{l = 1}^{k^\prime - 1} (1 - \pi_{l}), \label{eq:stickbreak1}\\
 \pi_{k^\prime} & \iid {\rm Beta}(1, \alpha). \label{eq:stickbreak2}
\end{align}
	   
Equations~\eqref{eq:clusterassign},~\eqref{eq:stickbreak1},~and~\eqref{eq:stickbreak2} are the key to understanding how the Dirichlet process mixture makes non-parametric estimation possible.
At the first step in the data generating process, we assign each observation to one of clusters $k^\prime = 1, 2, \dots$.
The assignment probabilities are determined by equations~\eqref{eq:stickbreak1}~and~\eqref{eq:stickbreak2}, which is called the ``stick-breaking'' process.
The origin of the name sheds light on how this process works.
When deciding the probability of the first cluster ($k^\prime = 1$), a stick of length $1$ is broken at the location determined by the Beta random variable ($\pi_1$).
The probability that each observation is assigned to the first cluster is set to be the length of the broken stick, $\pi_1$.
Next, we break the remaining stick of length $1 - \pi_1$ again at the place $\pi_2$ within the remaining stick.
The length of the second broken stick ($\pi_2 (1 - \pi_1)$) is used as the probability of each observation being assigned to the second cluster.
After setting the assignment probability of the second cluster, we continue to break the remaining stick following the same procedure an infinite number of times.
The probabilities produced by the stochastic process vanish as the cluster index increases because the remaining stick becomes shorter every time it is broken.
Although we do not fix the maximum number of clusters and allow the number to diverge in theory, the property of the stick-breaking process that causes the probability to quickly shrink towards zero prevents the number of clusters from diverging in practice.\footnote{The value of the prior parameter $\alpha$ determines how quickly the probabilities to form a new cluster vanish. For $\alpha = 1$, the Beta distribution in equation~\eqref{eq:stickbreak2} turns out to be the uniform distribution.
This is the standard choice in the literature (and is our default option in all results presented here), whereas a smaller (larger) value of $\alpha$ leads to a faster (slower) decrease in the cluster probabilities, depending on the total number of respondents in each cluster. Rather than experiment with defining different values for this hyper-parameter for problems of different sizes, we adopt a fully Bayesian approach and define an Gamma hyper-prior over $\alpha$,
 \[
 \alpha\sim \text{Gamma}(a_0,b_0)
 \]
 and learn a posterior distribution over $\alpha$ supported by the data. 
}

%We are now ready to apply these general principles to the commonly used 2-parameter IRT model, which yields the \emph{Multiple Policy Space} model we propose. 

%  \subsubsection*{The Multiple Policy Space Model}
% Let $y_{i,j}\in\{0,1\}$ be respondent $i$'s ($i\in{1,\ldots,N}$) response on item $j\in{1,\ldots,J}$. Our 2-parameter IRT model defines
% \begin{align}
% \begin{split}
%   y_{i,j} \mid \bm{\theta},\bm{\beta},\gamma &\iid \mathcal{B} \l( \Phi \l( \bm{\beta}_{k[i],j}\addtop \bm{\theta}_{i} - \gamma_{k[i],j} \r) \r),\; \forall i,j \\
%   \bm{\theta}_i &\iid \mathcal{N}_{D} \l( \bm{0},\bm{\Lambda}^{-1} \r), \; \forall i\\
%  (\bm{\beta}_{k_j,j},\gamma_{k_j})&\iid \mathcal{N}_{D+1} \l( \bm{0},\bm{\Omega}^{-1} \r),\;  \forall j
%  \end{split}
% \end{align}
%  where $k[i] \in 1, \ldots$ is a latent cluster to which respondent $i$ belongs; $\bm{\theta}_i$ is a vector of latent
%  respondent positions on $D$-dimensional space;
%  $\bm{\beta}_{k,j}$ is a vector of cluster-specific
%  item-discrimination parameters; $\gamma_{k,j}$ is a
%  cluster-specific item-difficulty parameter.\footnote{$\bm{\Lambda}$ and $\bm{\Omega}$ are prior precisions of ideal points and item parameters, respectively, with $\bm{\Lambda}\equiv\mathbf{I}_D$ for identification purposes.} Substantively, cluster-specific item parameters reflect the possibility that the IRF is shared by respondents belonging to the same group $k$.

Accordingly, when clusters over which DIF occurs are unobserved (both in membership and in number), we can rely on this probabilistic clustering
 process over a potentially infinite number of groups. 
 %One such model is the process defined in Equations~\eqref{eq:clusterassign},~\eqref{eq:stickbreak1},~and~\eqref{eq:stickbreak2} above, so that the probability that a respondent is sorted into cluster $k\addprime$, $\pi_{k\addprime}$, is assumed to be 
%  \[
%  k\addprime,\pi_{k\addprime} \iid \text{stick}(\alpha)
%  \]
%  and $\text{stick}(\cdot)$ denotes the DP stick-breaking process described in the previous section. 
 In this context, each cluster $k\addprime$ effectively defines a (potentially) different item response function, which in turn allows us to automatically sort observations into equivalence classes within which measurement invariance is expected to hold, without guaranteeing that observations sorted into \emph{different} clusters will be comparable. Hence, our model partitions respondents across a (potentially infinite) set of multiple policy spaces.

 In general, the substantive interpretation of estimated clusters needs to be approached cautiously. While our model is useful for identifying which respondents perceive a common latent space with each other, it will generally \emph{overestimate} the total number of actual (i.e., substantively distinct) clusters in the data \citep{kyung2009, WomackEtAl2014}.\footnote{In the context of DP \emph{mixtures}, this issue arises as a result of multiple components having very similar (though not exactly equal) item parameters. Accordingly, and in contrast to models that rely on DP priors to approximate arbitrary densities (as is the case for DP random-effects models), clusters in DP mixtures can be thought of a proper sub-clusters --- partitions that are nested within actual, substantive groupings in the data.} In the MPS model, multiple DP clusters can be thought of as being part of the same substantive group --- even if their corresponding item parameters are not exactly the same. What is more, this sub-clustering phenomenon can exacerbate known pathologies of mixture modeling and IRT modeling, such as \emph{label switching} (i.e., invariance with respect to component label permutations) and \emph{additive and multiplicative aliasing} (i.e., invariance with respect to affine transformations of item parameters and ideal points).  

Thus, even if all respondents actually belonged to the same cluster $k'$, we could estimate more than one cluster (denoted here as $k''$) with the other clusters recovering the transformed set of item parameters $\bm{\beta}_{k_{''r},j} = (\bm{\beta}_{k',j}\addtop K)$ (where $K$ is a arbitrary rotation matrix). However, we would still be able to see that clusters $k'$ and $k''$ were similar by examining the correlation between $\bm{\beta_{k'}}$ and $\bm{\beta_{k''}}$, as well as the patters of correlation between these and the item parameters associated with other clusters. When sub-clustering is an issue, two sub-clusters can be thought of as being part of the same substantive cluster if their items are highly correlated, or of they share similar correlation patterns with parameters in other sub-clusters.\footnote{Correlations, not being a proper metric, can violate the triangle inequality. Thus, high correlations between any two sets of item parameters do not always guarantee similar patterns of association to the parameters of other clusters.}

Having presented the details of our model, we now present the results of a Monte Carlo simulation that illustrates its ability to accurately partition respondents across clusters and recover the associated item parameters within each cluster.

\section{Monte Carlo Simulations}
\label{sec:montecarlosimulations}
As an initial test of our MPS model, we conduct a Monte Carlo simulation to test the ability of our model to correctly recover our parameters of interest. We simulate a data set in which $N=1000$ respondents provide responses to $J=200$ binary items. Respondents are randomly assigned to one of three separate clusters with probabilities 0.5, 0.2, and 0.3 respectively. In each cluster, respondent ability parameters and item difficult and discrimination parameters are all drawn from a standard normal distribution. For starting values, we use $k$-means clustering to generate initial cluster assignments, and principal components analysis on subsets of the data matrix defined by those cluster assignments for starting ability starting values. Item difficulty and discrimination starting values were generated for each cluster and item by running probit regressions of the observed data on the starting ability parameter values by cluster. We run 1,000 MCMC iterations, discarding the first 500 as burn-in, and keeping only the sample that produces the highest posterior density as the maximum \emph{a posteriori} (MAP) estimate of all parameters and latent variables, to avoid issues associated with label switching.\footnote{%
\label{fn:iteration}
An anonymous reviewer pointed out that it would be useful to obtain information about uncertainty by keeping MCMC iterations rather than using one iteration as the MAP estimate.
While we agree with this, there are some technical difficulties.
First, keeping cluster assignments for all MCMC iterations requires a large memory size, especially because recent data sets for ideal point estimation contain a massive number of respondents \citep{imai2016fast}.
Moreover, since our model is a mixture model, label switching across iterations may mislead uncertainty measures.
By contrast, since cluster assignments are discrete random variables, getting uncertainty might not provide much additional information.
However, we are aware of the possibility that for some summary statistics of cluster assignments, e.g., the minimum proportion of voters who are in the same cluster as the legislators, uncertainty can be computed as an interval on the continuous scale.
While we did not use such measures in our application, these could be of interest in other contexts.
}

\begin{table}[t]
\centering
\begin{tabular}{|p{.5\textwidth}| p{.166\textwidth} p{.166\textwidth} p{.166\textwidth}|}
  \hline
\multicolumn{1}{|c|}{}  & \multicolumn{3}{c|}{Simulated Cluster} \\
 \multicolumn{1}{|c|}{Estimated cluster} & \multicolumn{1}{c}{1} & \multicolumn{1}{c}{2} & \multicolumn{1}{c|}{3} \\ 
  \hline
 \multicolumn{1}{|c|}{1} & \multicolumn{1}{c}{\phantom{00}0} &  \multicolumn{1}{c}{\phantom{00}0} &  \multicolumn{1}{c|}{\phantom{0}74} \\ 
 \multicolumn{1}{|c|}{2} & \multicolumn{1}{c}{\phantom{00}0} & \multicolumn{1}{c}{110} &   \multicolumn{1}{c|}{\phantom{00}0} \\ 
 \multicolumn{1}{|c|}{3} & \multicolumn{1}{c}{\phantom{0}99} & \multicolumn{1}{c}{\phantom{00}0} &   \multicolumn{1}{c|}{\phantom{00}0} \\ 
 \multicolumn{1}{|c|}{4} & \multicolumn{1}{c}{\phantom{00}0} &  \multicolumn{1}{c}{\phantom{0}99} &   \multicolumn{1}{c|}{\phantom{00}0} \\ 
 \multicolumn{1}{|c|}{5} & \multicolumn{1}{c}{\phantom{00}0} &  \multicolumn{1}{c}{\phantom{00}0} &  \multicolumn{1}{c|}{\phantom{0}79} \\ 
 \multicolumn{1}{|c|}{6} & \multicolumn{1}{c}{\phantom{00}0} &  \multicolumn{1}{c}{\phantom{00}0} &  \multicolumn{1}{c|}{\phantom{0}63} \\ 
 \multicolumn{1}{|c|}{7} & \multicolumn{1}{c}{139} &  \multicolumn{1}{c}{\phantom{00}0} &   \multicolumn{1}{c|}{\phantom{00}0} \\ 
 \multicolumn{1}{|c|}{8} & \multicolumn{1}{c}{\phantom{00}0} &  \multicolumn{1}{c}{\phantom{0}93} &   \multicolumn{1}{c|}{\phantom{00}0} \\ 
 \multicolumn{1}{|c|}{9} & \multicolumn{1}{c}{118} &  \multicolumn{1}{c}{\phantom{00}0} &   \multicolumn{1}{c|}{\phantom{00}0} \\ 
 \multicolumn{1}{|c|}{10} & \multicolumn{1}{c}{126} &  \multicolumn{1}{c}{\phantom{00}0} &   \multicolumn{1}{c|}{\phantom{00}0} \\ 
   \hline
\end{tabular}
\caption{Simulated vs. Estimated Clusters, MPS model: The estimated clusters recover the simulated clusters, but the sub-clustering phenomenon results in multiple estimated versions of the same cluster. For example, estimated clusters 2 and 4 represent two different ways to identify the simulated cluster 2. }
\label{table:clusters}
\end{table}

Table~\ref{table:clusters} shows a cross-tabulation of the simulated vs. estimated cluster assignments. The estimation procedure is able to separate the simulated clusters well, in the sense that none of the estimated clusters span multiple simulated clusters. However, we see evidence of the sub-clustering phenomenon discussed earlier. Members of simulated cluster 1, for instance, were split into estimated clusters 3, 7, 9 and 10. Since members of simulated cluster 1 were all generated using the same item parameters, the four estimated clusters that partition them are effectively noisy affine transformations of each other. Thus, we expect that the four sets of estimated item parameters for clusters 3, 7, 9 and 10 will be correlated. Simulated clusters 2 and 3 are similarly split between multiple estimated clusters, and we could expect these parameters to be similarly correlated.

In a real-case application, of course, access to the true underlying cluster memberships is not available. And as we discussed earlier, Dirichlet process mixtures are ideal for capturing the \emph{distribution} of parameters by discretizing their support into an infinite number of sub-clusters. As a result, many of these Dirichlet sub-clusters may share very similar parameter values, effectively representing the same substantive groupings in terms of item functionings. Accordingly, using DP mixtures for diagnosing DIF requires a formal procedure for establishing which sub-clusters belong together by virtue of sharing similar item parameters, and which contain observations that truly differ in their item functionings.

The practical issue of establishing equivalence across groups can be approached from a number of perspectives. For example, researchers could employ pair-wise equivalence tests on the item parameters \citep[see, e.g.,][for illustrations in Political Science]{rainey2014, hartman_hidalgo2018}, being careful to account for the problems raised by conducting multiple comparisons (e.g., using a Bonferroni-style correction, or the Benjamini-Hochberg procedure to control the false discovery rate). Given the potentially large number of pairings, however, we rely on an alternative approach that studies the second and third order information contained in the item parameter correlation matrix. Specifically, we study the graph induced by correlations across entire vectors of estimated item parameters to reconstruct substantive clusters from the sub-clusters identified through the DP mixture, and encourage applied researchers to follow the same approach.

To do so, we treat correlations among times parameters as the adjacency matrix of a weighted, undirected graph defined on the set of sub-clusters. The problem of finding substantive clusters can then be cast as the problem of finding the optimal number of \emph{communities} of sub-clusters on this graph --- a problem for which a number of approximate solutions exist \citep[for a succinct review, see][]{sinclair2016}. 

\begin{figure}
    \centering
    \includegraphics[scale=0.5]{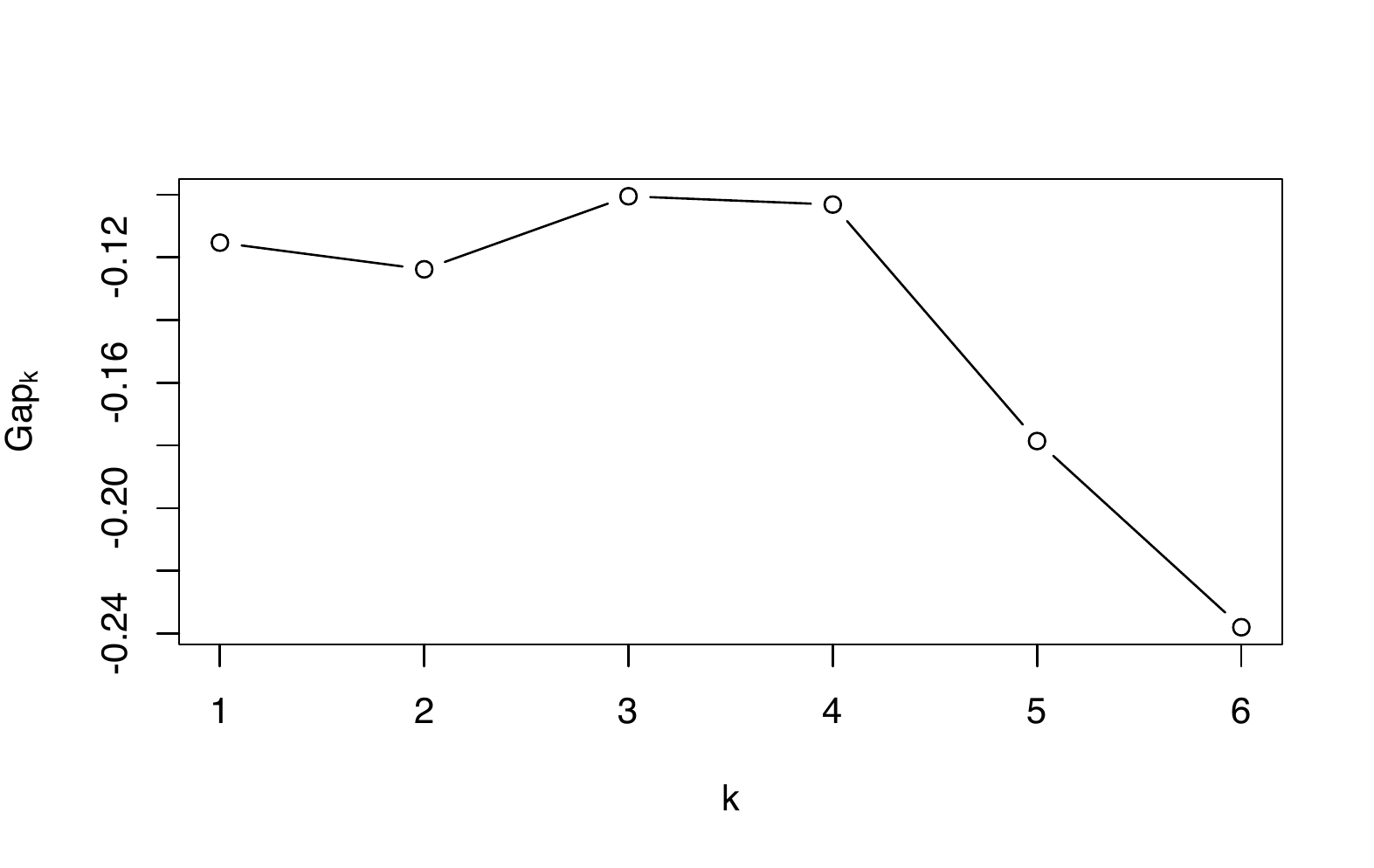}
    \caption{Gap statistic over different numbers of substantive clusters, defined as communities in a graph of item parameter correlations. High values of the gap statistic indicate a grouping with high within-cluster similarity relative to a null model (in which edges are drawn uniformly at random) with no heterogeneity. Thus, the $k$ that maximizes the gap statistic is a reasonable estimate for the number of substantive clusters in the data.}
    \label{fig:gap_het}
\end{figure}

For instance, a simple tool for identifying the optimal number of communities in a network is given by the \emph{Gap Statistic} \citep{tibshirani2001}, which compares an average measure of dissimilarity among community members to the dissimilarity that would be expected under a null distribution of edge weights emerging from a no-heterogeneity scenario:\footnote{Implementations can vary with respect to the way dissimilarity is operationalized and with respect to how the null distribution is defined.} 
\[
 \text{Gap}(k)=\E_{H_0}\left[\log(\bar{D}_k)\right] -\log(\bar{D}_k) 
\]
The optimal number of communities (i.e., of substantive clusters) can then be established by finding the $k^\star$ that maximizes $\text{Gap}(k)$. Figure~\ref{fig:gap_het} shows the value of gap statistic for different values of $k$, suggesting that the correct number of substantive clusters is 3 or 4.  

\begin{figure}[t]
    \centering
    \includegraphics[scale=0.4]{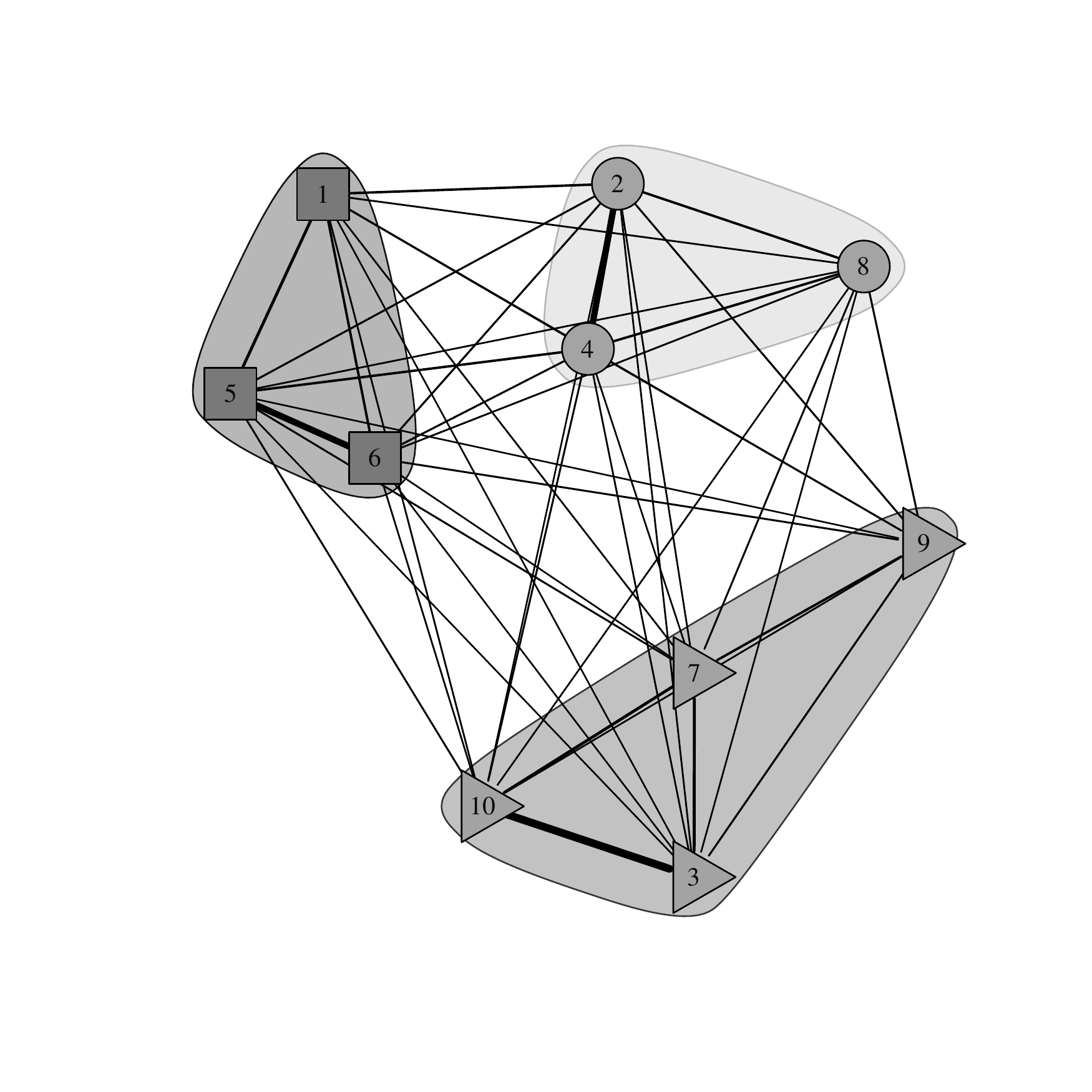}
    \includegraphics[scale=0.4]{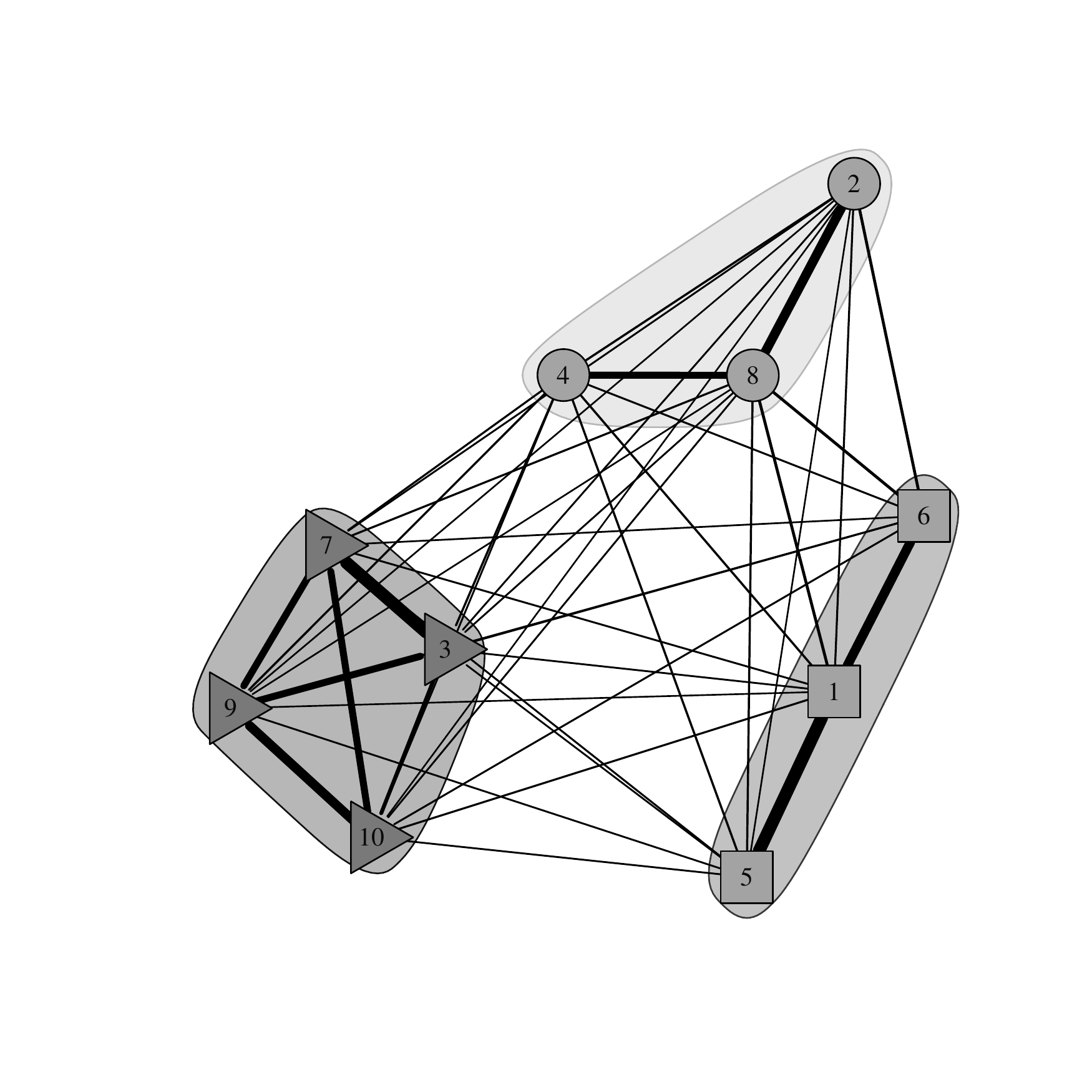}
    \caption{Graphs defined on nodes given by DP mixture sub-clusters: The graph has weighted edges defined using pair-wise correlations between discrimination parameters (left graph) and difficulty parameters (right graph). True simulation clusters are denoted with different node shapes, and communities detected by a modularity-maximizing algorithm are denoted with shaded regions. Recovery is of simulated clusters is exact in both instances.}
    \label{fig:graphSim}
\end{figure}

Indeed, Figure~\ref{fig:graphSim} shows the result of applying a simple community detection algorithm\footnote{Given the small number of sub-clusters in our estimation, we use a greedy procedure that starts by assigning each sub-cluster to its own community, and then proceeds to bind them together while locally optimizing a measure of \emph{modularity} --- the extent to which edge density is higher within communities than it is between them \citep{newman2003}.} to the graphs formed by using correlations across discriminations (left panel) and correlations across difficulties (right panel). In both instances, the true simulated clusters are denoted using shapes for the graph nodes, and the substantive groupings discovered by the community detection algorithm are denoted using shaded areas. In all instances, the communities identified map perfectly onto the known simulation clusters.      

\begin{figure}[!t]
  \begin{center}
  \includegraphics[scale=0.7]{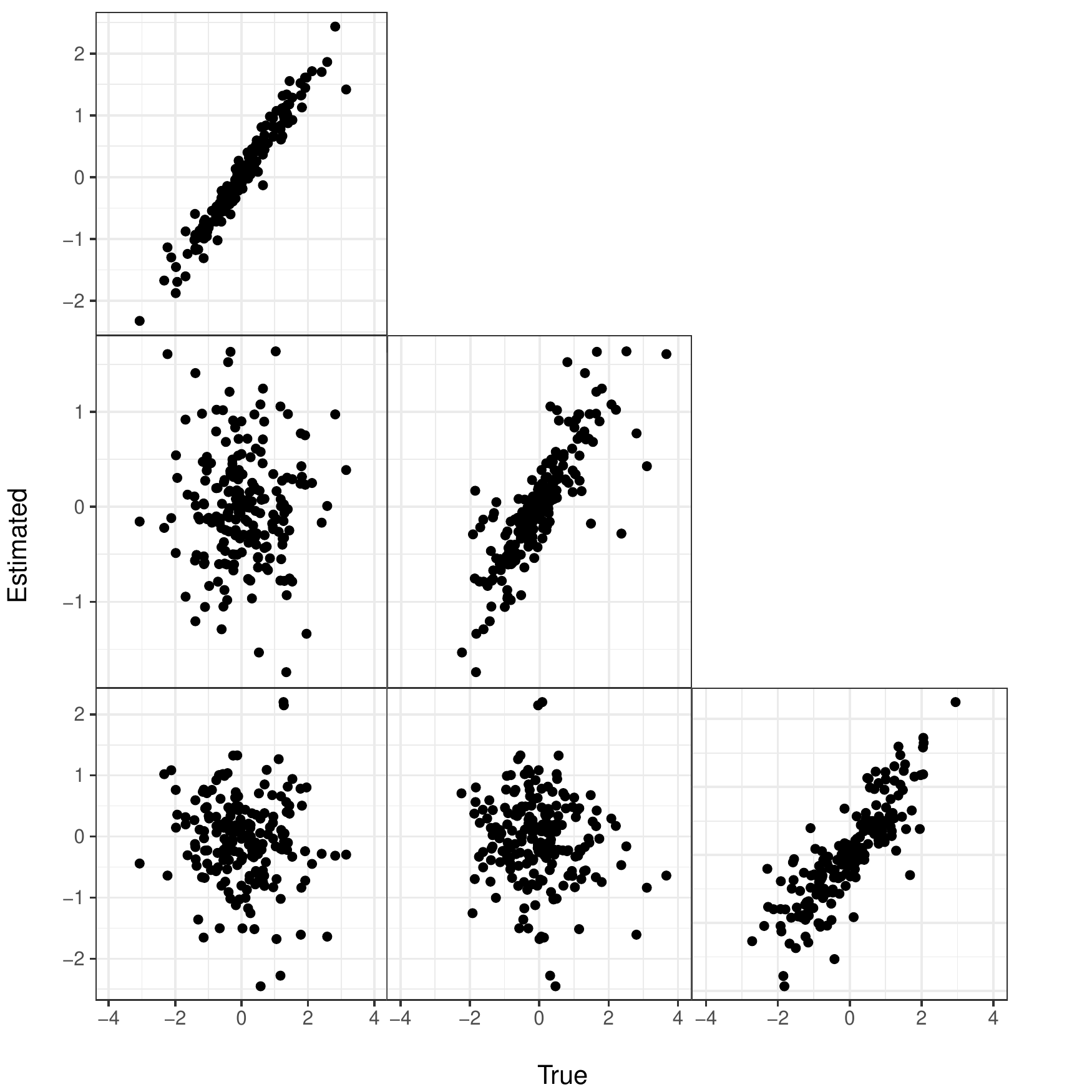}
\end{center}
\vspace{-.3in}
\caption{Correlation of Item Discrimination Parameters: Main diagonal plots estimated vs. simulated parameters for each cluster and show that the item discrimination parameters are correctly recovered to an affine transformation. Off-diagonal plots show cross-cluster correlation between estimated and true item parameters, which is expected (under the simulation) to be zero.}
\label{fig:item}
\end{figure}

While our previous analyses tested the correspondence between the true and estimated clusters, they say little about the recovery of the correct item parameters. In Figure~\ref{fig:item}, we explore the item discrimination parameters in a series of plots, where each panel plots two sets of item discrimination parameters against each other. Along the main diagonal, we plot combinations of the simulated item discrimination parameters (columns) for each cluster against the estimated parameters (rows) for the corresponding known cluster. In all three cases, the item parameters are well recovered and estimates are highly correlated with truth, with correlations of $r = 0.99$, $r = 0.97$, and $r = 0.97$ for the three plots.\footnote{In all cases, and because of the identification problems discussed earlier, estimates are only identified to an affine transformation of the true parameters. We therefore rotate all estimated parameters so that they match their known signs under the correspondence in Table~\ref{table:clusters}.} 

In turn, the off-diagonal terms present each combination of the \emph{simulated} item discrimination parameters vs. their (mis-matched) counterparts in other clusters. Since parameters in each cluster were generated from independent draws, the items are uncorrelated in reality. As expected, this independence is reflected in the estimated item parameters, which appear similarly uncorrelated with one another and with parameters in other known clusters.  

\begin{figure}[!t]
%  \spacingset{1}
%  \vspace{1.25in}
  \begin{center}
  \includegraphics[scale=0.7]{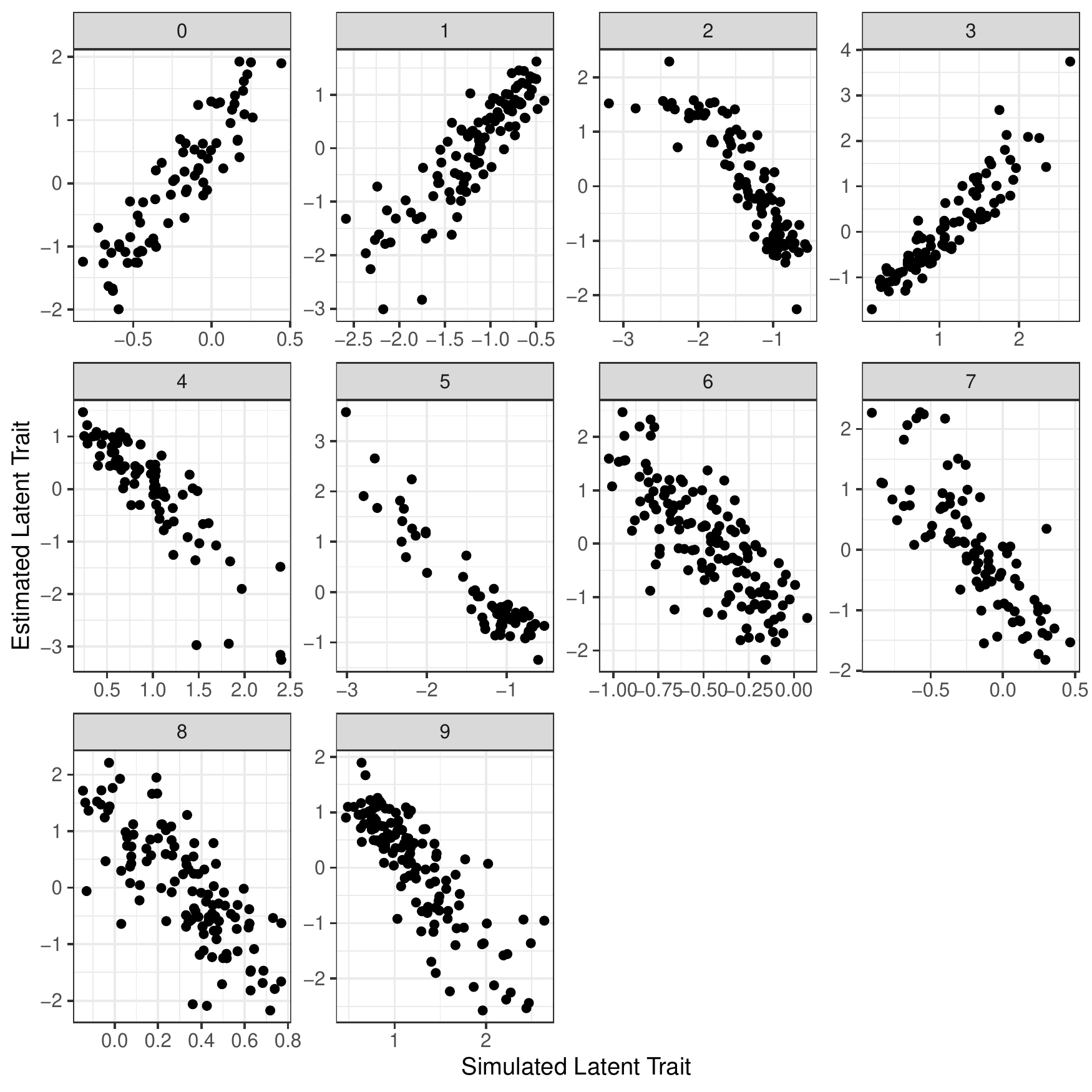}
\end{center}
\vspace{-.3in}
\caption{Correlation of Latent Traits Parameters: Plots show simulated against estimated latent traits for all 10 estimated clusters.}
\label{fig:traits}
\end{figure}

We repeat the same exercise in Figure~\ref{fig:traits}, but this time for the latent traits. In all cases, the latent traits are highly correlated, again demonstrating correct recovery of the traits of interest. The figures also highlight the fact that, in the MPS model, estimated latent traits are only comparable to other respondents belonging to the same cluster. If the MPS model facilitated comparisons across clusters, then at a minimum all of the figures shown here would consistently either be positively or negatively correlated with the simulated true ideal point. However, this is not the case. This is of course not surprising --- the MPS model effectively estimates a separate two-parameter IRT model for each cluster of legislators, allowing the same items to assume different item parameters for each group.  Thus, ideal points across groups would not be comparable, any more than ideal points from separate IRT models would be comparable. Of course, the MPS model makes a significant innovation in this regard --- it allows us to use the data itself to sort respondents into clusters, rather than forcing the researcher to split the sample \emph{a priori}.

Notably, standard measures of model fit also suggests that the MPS model fits the data better in the Monte Carlo. The MPS model produced a log-likelihood of $-85,776.71$, but when we fit the standard IRT model on the data that constrains all legislators to share the same single cluster, the log-likelihood drops significantly to $-117,477.2$. This improvement in fit is not surprising --- compared to standard 2P-IRT, MPS fits a much more flexible model. Whereas the standard, single cluster model involves estimating 1,000 respondent and 400 item parameters for a total of 1,400 parameters, the MPS model estimates 1,000 respondent parameters and 400 item parameters \emph{per cluster}. Since the maximum number of clusters in the estimation is set to 10, effectively the MPS model estimates 5,000 total parameters. Thus, a better measure of fit would penalize MPS for the added flexibility afforded by the substantial increase in parameters. The Bayesian Information Criterion (BIC) offers one such measure. It is equal to 252,043 for the single cluster model and for 232,604.7 the MPS model, which confirms that the MPS model fits the data better --- even after accounting for the substantial increase in model flexibility. Note that this BIC test is essentially a test of DIF across the identified clusters using methods similar in spirit to those proposed by \citet{lord2012applications} and \citet{thissen1993detection}.

\begin{figure}[t!]
    \centering
    \includegraphics[scale=0.5]{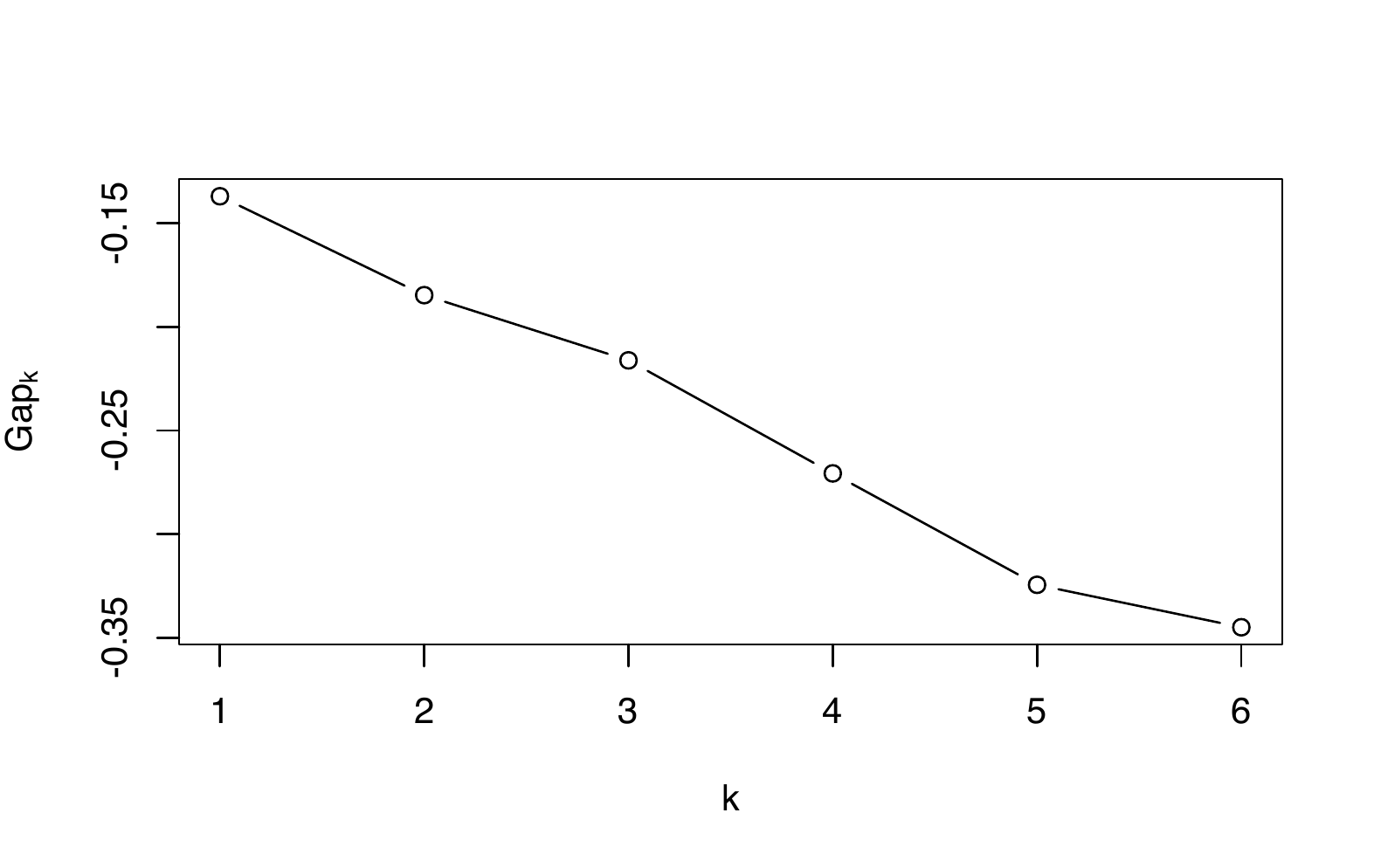}
    \caption{Gap statistic: Statistic defined over different numbers of substantive clusters, when true DGP has no heterogeneity. In this case, the gap statistic again recommends the correct number of clusters --- one, in this case.}
    \label{fig:gap_nohet}
\end{figure}

Finally, it is important to note that while MPS will partition observations into sub-clusters even when there is no underlying heterogeneity (i.e., even when the standard IRT model is correct), the similarity of item parameters across sub-clusters will immediately suggest that the resulting partition is substantively spurious. To see this, consider Figure~\ref{fig:gap_nohet}, which depicts the values of the gap statistic as computed on a graph defined as those in Figure~\ref{fig:item}, but resulting from a model estimated on data that has no underlying heterogeneity in IRFs. The gap statistic correctly suggests that the correct number of substantive clusters is, in fact, 1. The idea that there is no heterogeneity is further supported by the fact, under such a data-generating process, the standard IRT model with a single cluster fits the data better, with $\text{BIC}_{\text{IDEAL}}= 168430.8$ versus $\text{BIC}_{\text{MPS}}=173686.3$. Thus, there is little evidence that MPS will overfit data when there is no heterogeneity to be identified.

We now turn to our original motivating application: evaluating whether (or rather \emph{which}) U.S. voters  can be scaled on the same space as their legislators.

\section{Empirical Results}

We apply the MPS model to one of the main examples used in \citet{jessee2016can} --- the 2008 Cooperative Congressional Election Study (CCES). This is an online sample of 32,8000 survey respondents from the YouGov/Polimetrix panel administered during October and November 2008. In total, the CCES included eight bridging items that directly corresponded to votes taken during the 110th House and Senate, which can be matched to 550 legislators.\footnote{We lose 2 legislators who recorded no votes on any of the items under study.} The policy items included withdrawing troops from Iraq within 180 days, increasing the minimum wage, federal funding of stem cell research, warrantless eavesdropping of terrorist suspects, health insurance for low earners, foreclosure assistance, extension of free trade to Peru and Columbia, and the 2008 bank bailout bill.\footnote{%
\label{fn:commonitemsassumption}
The example here makes the same assumption that all joint scaling papers make --- that legislators and voters understand the roll call item in a consistent manner. This is known to not literally be true --- see \citet{hill2019meaning} for research on how legislators and voters may understand even the same roll call vote differently. However, for the purpose of detecting DIF with our model, we adopt and focus on the ``common understanding of items'' assumption that is prevalent throughout this literature.} In this example, Jessee found that joint scaling appeared to work relatively well for this data set --- that is, the ideal points from the grouped model look relatively similar regardless of whether one uses item parameters derived from respondents, the House, or the Senate.

We run 110,000 MCMC iterations, discarding the first 10,000 as burn-in, and keeping only the MAP estimate of the parameters of interest. The maximum number of clusters is constrained to be 10. Similar to the Monte Carlo, we generate starting ideal point values using principal components analysis within each cluster, and probit regression for starting item parameter values. However, rather than generating initial cluster assignments using $k$-means clustering, we instead start all legislators in one cluster, and all voters in a second cluster. Legislators are constrained to remain in the same cluster throughout each iteration, but voters are permitted to change cluster memberships.\footnote{%
\label{fn:constraint}
This constraint fits the substantive question (i.e., identifying which voters move into a cluster occupied by all legislators), but we acknowledge that for other substantive questions, it may be appropriate to set other constraints. For example, one could separate Southern and Northern Democrat legislators into separate fixed clusters and allow voters to move into those clusters.} 

Table 2 shows a cross-tabulation of the final estimated clusters on the rows against the two separate starting clusters for the legislators and voters. All 550 legislators start in the same cluster, and are constrained to remain so (although their ideal points within the cluster are permitted to change). In turn, the 32,800 surveyed voters divide themselves across 6 different clusters, with 15,732 respondents remaining in the same cluster as the legislators.

\begin{table}[t]
\centering
\begin{tabular}{| c | c c |}
  \hline
Estimated Cluster & Legislator Starting Cluster & Voter Starting Cluster \\ 
  \hline
1 & 550 & 15732 \\ 
  2 &   0 & 8256 \\ 
  3 &   0 & 7469 \\ 
  4 &   0 &  17 \\ 
  5 &   0 & 114 \\ 
  6 &   0 & 964 \\ 
   \hline
\end{tabular}
\caption{Estimated vs. Starting Clusters: Legislators all started in cluster 1, and remained there throughout estimation.}
\label{table:CCES_cluster}
\end{table}

The 15,732 respondents estimated to share the same cluster with the legislators are almost certainly underestimated, due to the fact that different clusters in DP-prior models may nevertheless share similar parameter values. Table~\ref{tab:CCES_correl} explores this further, tabulating the correlations of the item discrimination parameters between each of the 6 populated estimated clusters. From examining this table, we see that estimated clusters 2 and 5 have item parameters than are highly correlated with those in the constrained legislator cluster. Combining respondents from clusters 1, 2, and 5 together, 24,102 of the 32,800 respondents in the CCES sample, or approximately 73\% of the sample, lie in the same ideological space as legislators.

\begin{table}[t]
\centering 
%\begin{tabular}{@{\extracolsep{5pt}} c | d{2.3}d{2.3}d{2.3}d{2.3}d{2.3}d{2.3}} 
\begin{tabular}{@{\extracolsep{5pt}} c | cccccc} 
%\\[-1.8ex]
\hline 
\hline \\[-1.8ex] 
   & \multicolumn{6}{c}{Estimated Cluster} \\
Estimated Cluster & 1 & 2 & 3 & 4 & 5 & 6  \\ 
\hline \\[-1.8ex] 
1 &  &  &  &  &  &  \\ 
  2 & .76 (.27) &  &  &  &  &  \\ 
  3 & -.43 (.37) & -.14 (.40) &  &  &  &  \\ 
  4 & .13 (.40) & -.10 (.41) & -.80 (.25) &  &  &  \\ 
  5 & -.75 (.27) & -.62 (.32) & .37 (.38) & -.41 (.37) &  &  \\ 
  6 & -.13 (.40) & -.00 (.41) & -.49 (.36) & .32 (.39) & .33 (.39) &  \\ 
\hline
%   \\[-1.8ex] 
\end{tabular}
\caption{Correlations of Item Discrimination Parameters between Estimated CCES 2008 Clusters: Standard errors in parenthesis} 
   \label{tab:CCES_correl} 
\end{table}

With this large number of observations falling in a single cluster, it is not surprising that different model selection criteria provide different indications as to whether a standard IRT or MPS fits the data better. For instance, while the comparison between the BIC produced by our model (viz., 408,016.4) and the BIC produced by a standard IRT model (viz., 407,033.7) would suggest the latter offers a better fit to these data, the evidence is reversed when we consider AIC as a selection criterion (with values of 355,419.4 and 370,214.8 for MPS and the regular IRT, respectively.). Nevertheless, an evaluation of the extent to which communities of sub-clusters emerge from these pairwise correlations suggests the importance of separating between two sets of voters.

\begin{figure}[t!]
    \centering
    \begin{minipage}{6in}
  \centering
  \raisebox{-0.5\height}{\includegraphics[height=2.0in, scale=0.7]{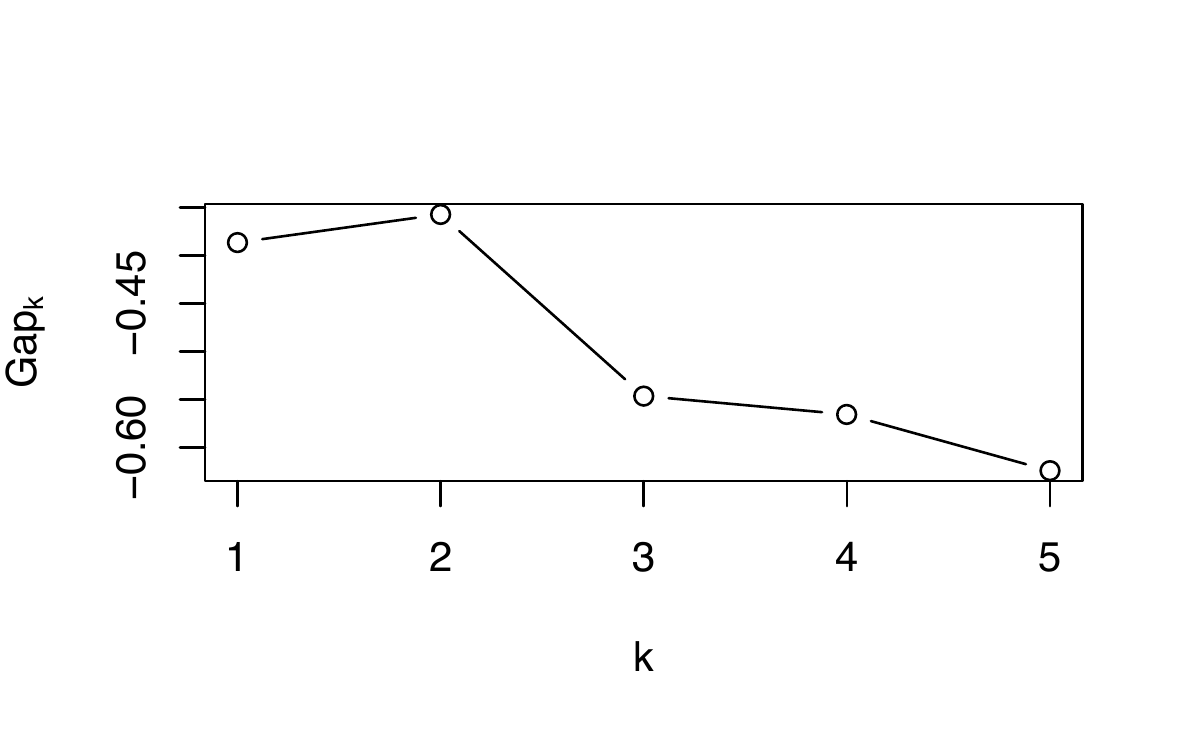}}
  \hspace*{-.7in}
  \raisebox{-0.5\height}{\includegraphics[height=3.2in, scale=1.2]{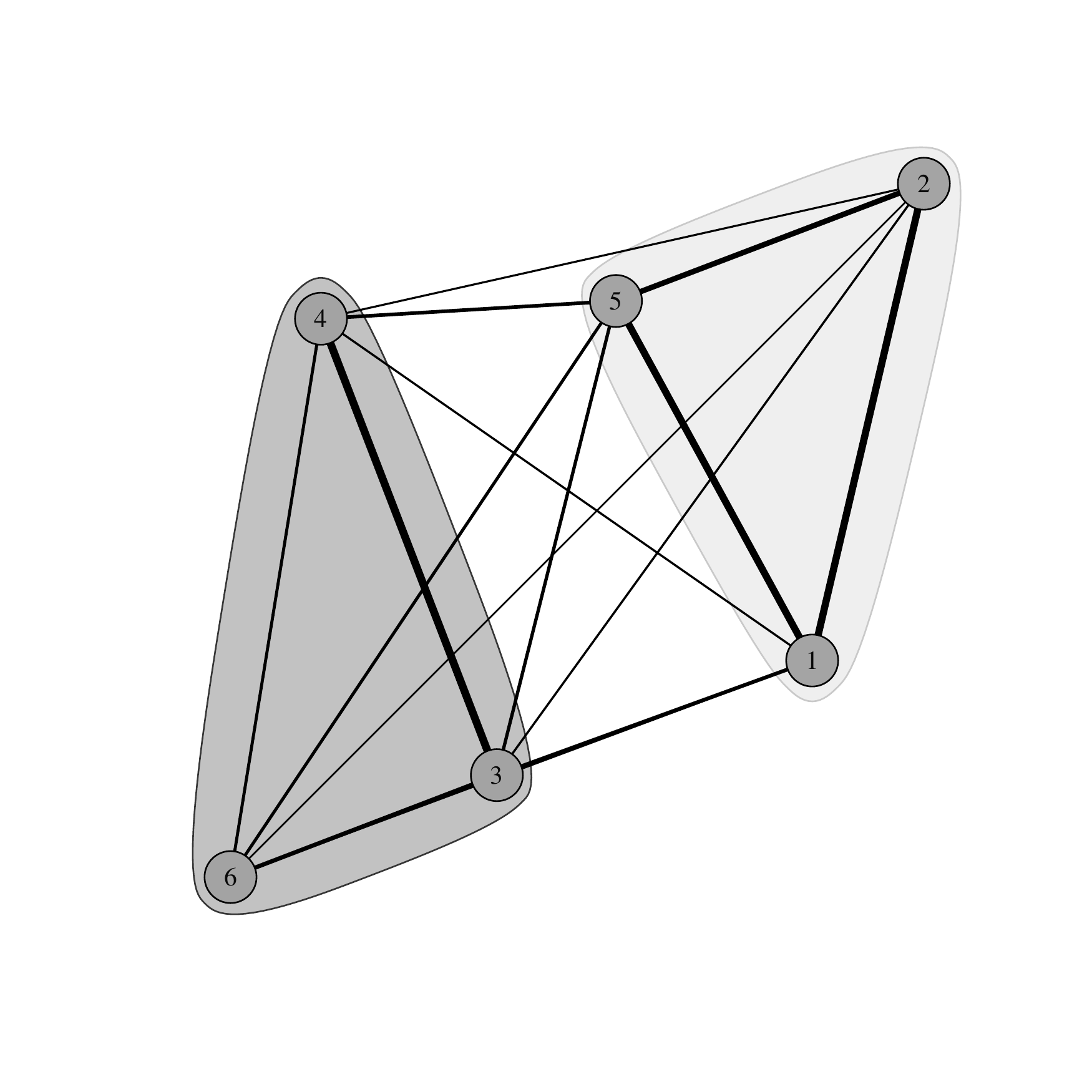}}
\end{minipage}
    \caption{Left Panel: Gap statistic; Right Panel: Graph on nodes given by DP mixture sub-cluster. Left panel shows two substantive clusters appear to fit the data best. Right panel graph has weighted edges defined using pair-wise correlations between discrimination parameters in a model estimated on the 2008 CCES data. Shaded regions denote communities detected by a modularity-maximizing algorithm. Again, two substantive clusters appear summarize the data best, with a ``legislator cluster'' formed by sub-clusters 1, 2, and 5.}
    \label{fig:corrJessee}
\end{figure}

The right panel of Figure~\ref{fig:corrJessee} depicts this correlation-weighted graph, along with the substantive clusters identified by the same greedy algorithm used in the previous section (indicated using gray shaded areas). In this case, both the greedy community-detection procedure and the gap statistic (depicted on the left panel of Figure~\ref{fig:corrJessee}) identify two communities --- one containing all legislators and a large number of voters, and another composed of the remaining voters who \emph{do not} share the same policy space as legislators.      

To further validate this sorting, we study the extent to which a model that forces all voters in sub-clusters 1, 2, and 5 to remain fixed in the cluster containing all legislators, and evaluate whether such a model results in a better fit to the observed responses. Such a model results in an unequivocally better fit vs. a model that allows all voters to be freely allocated to clusters, with a BIC of 407,426.8 and an AIC of 365,820.8.\footnote{In turn, a model that only fixes the membership of the 15,732 voters who are estimated to be in cluster 1 results in a BIC of 407,623.5 and an AIC of 368,304.6, again indicating a worse fit than a model in which everyone in clusters 1, 2, 5 are fixed from the beginning.}

\begin{figure}[t]
\centering
\includegraphics[scale=0.7]{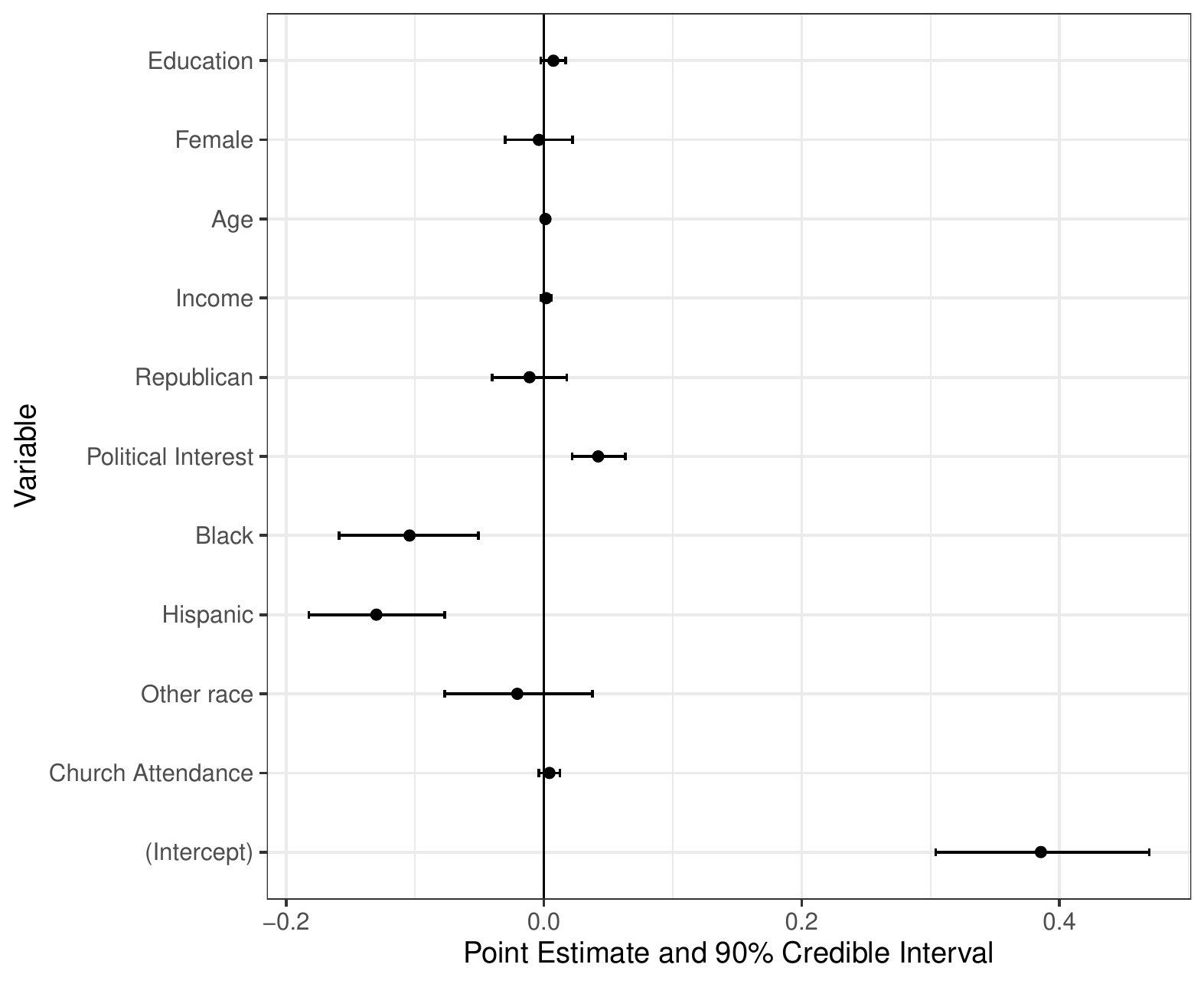}
\caption{Point estimates and 90\% credible intervals for coefficients in Bayesian probit regression of Membership into Estimated Legislator Cluster. A reference line is added at zero. We find that `Political Interest', `Race', and `Age' are likely to be characteristic of voters in the legislator cluster.} 
\label{fig:legis} 
\end{figure}

In addition, and to explore the question of what characterizes the 24,102 survey respondents who ``think like a legislator'' (i.e., who are sorted into estimated clusters 1, 2, and 5), we group these respondents together and predict membership in this pseudo-legislator group with a Bayesian binomial probit regression (with vague, uniform priors), using a range of standard covariates --- including education, gender, age, income, race, party identification, political interest, and church attendance. We report these results in Figure~\ref{fig:legis}.\footnote{We fit our model using R function \texttt{MCMCPack::BayesProbit()}, in package version 1.6-3. We take 9,000 samples from the posterior, having discarded the first 1,000 samples as burn-in.} 

We find that older voters and people who express more interest in politics all tend to map their latent traits onto observed responses similarly to the way legislators do, while Black and Hispanic voters are less likely than their white counterparts to share an ideological space with legislators. And while the coefficients associated with education, income and gender all fail to attain our chosen level of significance, their signs do indicate that more educated and richer voters also tend to think more like legislators, while women appear less likely to share the policy space of their (mostly male) legislative counterparts.

Overall, our findings are largely consistent with Jessee, who found that latent trait estimates from this data set were consistent regardless of whether one used the item parameters estimated from legislators or voters. However, the key difference from our approach is that we not only identify the 73\% of survey respondents who follow this pattern, but also the 27\% of survey respondents that do not share an ideological space with legislators. Furthermore, our improved fit statistics suggests that the improvement in model fit for this subset of respondents is quite significant, even for a data set where the recovered ideal points would be somewhat similar regardless of whether one used only the voter, House, or Senate item parameters to generate ideal points.  

% \section*{Confirm these reproductions}

% \begin{itemize}
% \item Monte Carlo simulation
% \begin{itemize}
%     \item Simulated vs Estimated cluster, Table 1
%     \item Tabulated correlations of Item parameters, Table 2
%     \item Scatterplots of Item parameters, Figure 1
%     \item Correlation of latent trait parameters, Figure 2. Numeric correlations pg. 15
%     \item Log likelihood and AIC in text (pg. 15)
% \end{itemize}

% \item Jessee legislator example
% \begin{itemize}
%     \item Simulated vs Starting cluster, Table 3
%     \item Tabulated correlations of Item parameters, Table 4
%     \item probit regression, Table 5
%     \item AIC comparison pg. 18
% \end{itemize}
%\end{itemize}

\section{Conclusion}
When implementing commonly used measurement models, most researchers implicitly subscribe to the idea that all individuals share a common understanding of how their latent traits map onto the set of observed responses: legislators are believed to have shared sense of where the cut-point between voting alternatives lies, survey respondents are assumed to ascribe a common meaning to the scales presented in the questions they confront, and voters are understood to perceive the same candidates and parties as taking on similar ideological positions.    

When this assumption is violated by the real data-generating process, however, adopting this widespread strategy can be a costly over-simplification that results in invalid measures of the characteristics of interest. By assuming units can be separated into groups for whom comparable item functioning holds, we propose a modeling strategy that relaxes the stringent measurement invariance assumption, allowing researchers to identify sets of incomparable units who can be mapped onto multiple latent spaces. The distinctive feature of our proposed approach is that it does not require \emph{a priori} identification of group memberships --- or even a prior specification of  the number of heterogeneous groups present in the sample.

On this note, it is important to reiterate that the clusters we obtain from our Dirichlet Process prior models are not distinct groups, in the sense that they may share parameters that are similar enough to be considered part of the same sub-population. Our models, therefore, are designed to account for the existence of these heterogeneous groups without directly identifying \emph{a posteriori} memberships into them. In so doing, our models assume the target of inference is the latent traits, rather than the group memberships. And while it is sometimes possible to tease out sub-populations from estimated Dirichlet Process clusters, we generally discourage users from trying to ascribe substantive meaning to the clusters directly identified by our non-parametric model --- except to say that observations that are estimated to be in the same Dirichlet Process cluster have latent traits that can be safely compared to one another. If a more thorough interpretation of which sub-clusters are, in fact, substantively equivalent is of interest, we encourage researchers to post-process the Dirichlet mixture clusters in order to identify the more substantive groupings defined by item parameters that are similar enough, as we did through the use of the Gap statistic on the graph of item parameter correlations in our illustration of the MPS model.\footnote{%
\label{fn:similarenough}
The meaning of ``similar enough'' is, of course, a matter of researcher discretion. In our illustration, we relied on the gap statistic and community detection tools defined on the correlation graph of item discriminations. Alternative approaches that make the notion of sufficient similarity more explicit could rely on equivalence tests, as they require the definition of a clear equivalence range.} Having done so, researchers can then make data-driven decisions about the presence and pervasiveness of differential item functioning in their data. Alternatively, design-based solutions (such as anchoring vignettes) can help ascribe meaning to (different sub-groups), while other model-based approaches --- such as the product partition DP prior model proposed by \citet{WomackEtAl2014}, or the repulsive DP-mixture model proposed by \citet{xie2020} --- may offer potential analytical avenues, if adapted to the IRT framework. We leave these possibilities for future research.

Despite these caveats, we believe our proposed model can offer researchers a simple alternative to the standard modeling approach and its strong invariance assumptions.  If heterogeneity in item functioning is a possibility---as we suspect is often the case in the social science contexts in which probabilistic measurement tools are usually deployed---our approach offers applied researchers the opportunity to assess that possibility and identify differences across units if said differences are supported by the data, rather than simply assuming those differences across sub-populations away.

A broader substantive question that this paper does not address directly is whether our empirical results hold for joint scaling of legislators and voters using different data sets and/or in other contexts.
While we found that most voters share an ideological space with legislators in the CCES data set, it is still an open question whether most voters and legislators can be jointly scaled particularly when there are a greater number of bridging items that provide more information about how similar their item response functions are.
Having presented the methodology that allows researchers to address this question, we leave it for future research.

\clearpage
\singlespacing
\pdfbookmark[1]{References}{References}
\bibliography{refs}

\clearpage

\appendix
\section{Computational Details}
\paragraph{Gibbs Sampler.} Truncate the stick-breaking process at some constant $K$. Define 
  \begin{enumerate}
  \item Update the stick-breaking weight $\pi_{k^\prime}$ for $k^\prime = 1, \dots, K - 1$ by sampling from a Beta distribution s.t.
    \begin{align*}
      \pi_{k^\prime} \sim {\rm Beta} \l(1 + N_{k^\prime}, \alpha + \sum_{l = k^\prime + 1}^{K} N_l \r)
    \end{align*}
    where $N_k$ is the number of observations assigned to cluster $k$ under the current state.
  \item Update $k[i] \in \{1, \dots, K \}$ for $i = 1, \dots, N$ by multinomial sampling with
    \begin{align*}
      {\rm Pr}(k[i] = k^\prime \mid \bm{y}_{i}, \, \bm{\theta},\bm{\beta}, \bm{\gamma} ) \propto p_{k^\prime} \, {\rm Pr}\l( \bm{y}_{i} \mid \bm{\theta}_i,\bm{\beta}_{k^\prime},\bm{\gamma}_{k^\prime} \r)
    \end{align*}
    where
    \begin{align*}
      p_{k^\prime} &\equiv \pi_{k^\prime} \prod_{l = 1}^{k^\prime - 1} (1 - \pi_{l}) \\
      {\rm Pr}\l( \bm{y}_{i} \mid \bm{\theta}_i,\bm{\beta}_{k^\prime},\bm{\gamma}_{k^\prime} \r)
                   & = \l( \Phi \l( \bm{\beta}_{k\addprime,j} \bm{\theta}_{i} - \gamma_{k\addprime,j} \r) \r)^{y_{ij}}
                     \l( 1 - \Phi \l( \bm{\beta}_{k\addprime,j}\bm{\theta}_{i} - \gamma_{k\addprime,j} \r) \r)^{1 - y_{ij}}
    \end{align*}
    In practice, we augment the latent variable $y_{i,j}^\ast$ so that we have:
\[
      {\rm Pr}(k[i] = k^\prime \mid \bm{y}_{i}\addast, \, \bm{\theta}_i,\bm{\beta}_{k\addprime}, \bm{\gamma}_{k\addprime} )
      \propto
      p_{k^\prime} \,   \mathcal{N}\l( y_{i,j}^\ast \mid \bm{\beta}_{k\addprime, j}\addtop \bm{\theta}_i - \gamma_{k\addprime , j} , \, 1 \r)
    \]

  \item Conditional on $\boldsymbol{\theta}$, $\boldsymbol{\beta}$, $\boldsymbol{\gamma}$ and $\bm{k}$, sample
    \[
      y_{i,j}^\ast \sim
      \begin{cases}
        \mathcal{N}(\theta_i\beta_{k^\prime, j} - \gamma_{k^\prime, j}, 1)\mathcal{I}(y_{i,j}^\ast < 0) &\text{if $y_{i,j}=0$}\\
        \mathcal{N}(\theta_i\beta_{k^\prime, j} - \gamma_{k^\prime, j}, 1)\mathcal{I}(y_{i,j}^\ast \geq 0) &\text{if $y_{i,j}=1$}
      \end{cases}
    \]
    which can be parallelized over respondents and items, for dramatic speedups.
  \item Conditional on $\bm{\theta}$, $\bm{y}^\ast$ and $\bm{k}$, sample
    \[
      (\bm{\beta}_{k^\prime,j}, \gamma_{k^\prime,j}) \sim \mathcal{N}_{D+1}\left(\bm{\mu}_{k^\prime,j},\bm{M}_{k^\prime,j}^{-1}\right)
    \]
    where $\bm{M}_{k^\prime, j}=(\bm{X}_{k^\prime}^\top\bm{X}_{k^\prime}+\bm{\Omega})$; $\bm{\mu}_{k^\prime,j}=\bm{M}_{k^\prime, j}^{-1}\bm{X}_{k^\prime}^{\top}\bm{y}^\ast_{k^\prime,j}$; $\bm{X}_{k^\prime}$ is a matrix with typical row given by $\bm{x}_i=[\bm{\theta}_i,-1]$ for $i$ s.t. $k[i]=k^\prime$, and $\bm{y}^\ast_{k^\prime,j}$ is a vector with typical element $y^\ast_{i,j}$, again restricted to $i$ s.t. $k[i]=k^\prime$. 
    
    Once again, this can be parallelized over items and clusters, reducing user computation times. 
  \item Conditional on $\bm{\beta}$, $\bm{\gamma}$ and $\bm{k}$, and for each $i$ s.t. $k[i]=k^\prime$, sample
    \[
      \boldsymbol{\theta}_i \sim \mathcal{N}_{D}(\bm{\nu}_{k^\prime}, \bm{N}_{k^\prime}^{-1})
      \]
    where $\bm{N}_{k^\prime}=\left(\bm{B}_{k^\prime}^\top\bm{B}_{k^\prime} + \bm{\Lambda}\right)$; $\bm{\nu}_{k^\prime}=\bm{N}_{k^\prime}^{-1}\bm{B}_{k^\prime}^\top\mathbf{w}_i$; $\bm{B}_{k^\prime}=[\bm{\beta}_{k^\prime,1},\ldots,\bm{\beta}_{k^\prime,J}]^\top$ is an $J\times D$ matrix, and $\bm{w}_i=\bm{y}^\ast_{i}+\bm{\gamma}_{k^\prime}$ is a $J\times 1$ vector.  We parallelize these computations over respondents. 
    \item Finally, conditional on cluster assignments and stick-breaking weights, sample
    \[
    \alpha \sim \text{Gamma}(a_0 + N - 1, b_0 - \sum_{k\addprime=1}^{N-1}\log(1-\pi_{k\addprime}))
    \]
  \end{enumerate}

\end{document}